**Brightness Characteristics of the Qianfan Satellites and Evidence That Some Are Tumbling**

Anthony Mallama*[1], Richard E. Cole[1], Stephan Hellmich[2], Roger Spinner[3], Jeff Warner[1] and Jay Respler[1]

[1] IAU - Centre for the Protection of Dark and Quiet Skies from Satellite Constellation Interference
[2] Laboratory of Astrophysics, École Polytechnique Fédérale de Lausanne (EPFL), Switzerland
[3] s2a systems

* Correspondence: anthony.mallama@gmail.com

2025 May 11

Abstract

The mean apparent magnitude of the Qianfan satellites is 5.76 +/- 0.04, while the mean of magnitudes adjusted to a distance of 1,000 km is 5.24 +/- 0.04, based on 1,161 observations. Light curves of several spacecraft display rapid periodic fluctuations which indicate that they are tumbling. Nearly all of the non-tumbling satellite observations can be modeled with diffusely reflecting, Earth-facing surfaces. The Qianfan constellation will impact astronomical research and aesthetic appreciation of the night sky unless their brightness is mitigated.

**1. Introduction**

Bright spacecraft interfere with observations of the night sky (Barentine et al. 2023 and Mallama and Young 2021). The Qianfan constellation is a special concern because 14,000 satellites are planned. A preliminary study of the 18 spacecraft of the first launch reported magnitudes ranging from 4 to 8 with brighter values occurring near zenith (Mallama et al. 2024).

New material reported in this paper includes comprehensive magnitude statistics that characterize spacecraft brightness, summaries of launches 2 through 5, a description of Qianfan orbit-raising, and evidence that some of the satellites are tumbling.

The contents are organized as follows. Section 2 describes the launches and orbits of Qianfan satellites. Section 3 describes how brightness measurements were obtained for this study. Section 4 characterizes the satellites' magnitudes from an empirical perspective.

Section 5 addresses spacecraft that appear to be tumbling. Section 6 explains the brightness characteristics of non-tumbling satellites using a physical model. Section 7 discusses the constellation's impact on astronomy and Section 8 presents our conclusions.

The Qianfan constellation is also known as Thousand Sails, G60 Starlink and by other names. To avoid confusion, individual spacecraft are identified herein by their NORAD numbers.

**2. Launches and orbits**

Until now, five launches of Qianfan satellites with 18 spacecraft apiece have occurred. The orbital inclinations are 89° and initial altitudes were near 800 km.

The first launch took place on 2024 August 6 and the satellites began orbit-raising a few weeks later. All but one of those spacecraft attained an altitude near 1,070 km. The exception, NORAD 60385, remained at its original height. That spacecraft has failed according to J. McDowell's website, https://planet4589.org. The satellites of Launch 1 were built by the Shanghai Microsatellite Engineering Center.

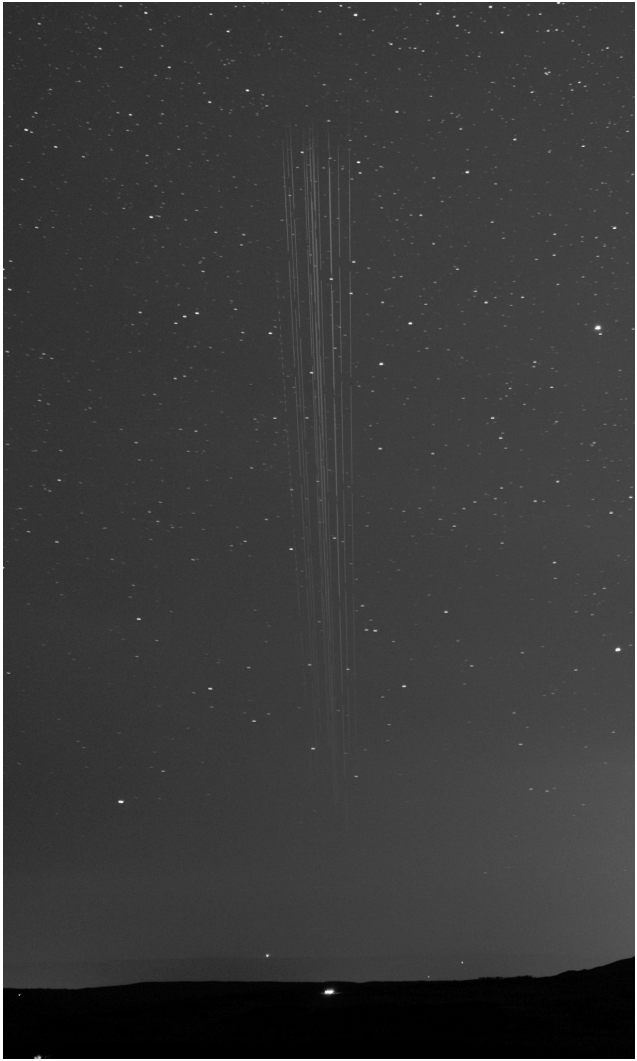

Figure 1. A composite image of the Qianfan constellation from Launch 1 recorded on 2024 September 3. Credit: catchingtime.com.

The spacecraft of Launch 2 were manufactured by a different company. These satellites, which were placed into orbit on 2024 October 15, have been less successful. McDowell's website indicates that nine of the satellites have failed and Figure 2 shows that many did not complete orbit-raising. Section 5 presents evidence that several of these failed spacecraft are tumbling.

Additional Qianfan launches have occurred on 2024 December 5, 2025 January 3 and 2025 March 11. Just one of those 54 satellites (NORAD 62240) has failed, according to McDowell.

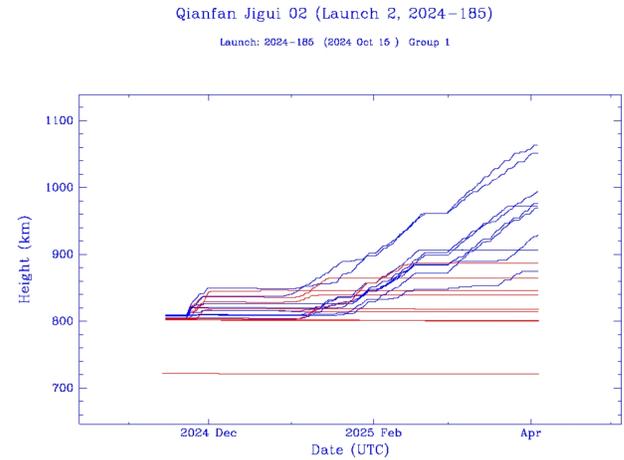

Figure 2. Orbiting-raising for Group 2. The red lines indicate satellites that have failed. Credit: J. McDowell.

### 3. Observations

Satellite magnitudes were determined using electronic and visual techniques. Some of the electronic measurements were obtained at the MMT9 robotic observatory (Karpov et al. 2015 and Beskin et al. 2017). MMT9 consists of nine 71 mm diameter f/1.2 lenses and 2160 x 2560 sCMOS sensors. The photometry is within 0.1 magnitude of the V-band based on information in a private communication from S. Karpov as discussed by Mallama (2021). We averaged the 10-Hertz data into 5-second intervals.

Electronic measurements of tumbling satellites were acquired using the CHE-1 and CHL-1 observatories, which are part of the global station network operated by s2a systems. The sensor at CHE-1 is an f/4.5 17" CDK (Corrected Dall-Kirkham) telescoped equipped with a C1X61000T CMOS camera, the sensor at CHL-1 is an f/2.2 14" RASA (Rowe-Ackermann Schmidt Astrograph) telescope equipped with a QHY600 CMOS camera. The exposure times used were 10 ms (CHE-1) and 100 ms (CHL-1).

Visual magnitudes were determined by comparing the spacecraft to nearby reference stars. The angular proximity between satellites and stellar objects accounts for variations in sky transparency and sky brightness. Mallama (2022) describes this method in more detail.

The brightness characterization in the following section is based on 1,161 V-band and visual magnitudes, of which 682 were obtained during *orbit-raising* and 98 others pertain to satellites that were tumbling or had otherwise failed. Five of the magnitudes are *specular,* meaning that sunlight was reflected in a mirror-like fashion from the satellite chassis directly to the observer which resulted in higher than normal brightness. The 376 magnitudes that are not orbit-raising, tumbling or specular are called *regular.* The data are available from the [SCORE](#) database maintained by the IAU/CPS.

### 4. Brightness characterization

Statistics of apparent magnitudes for Qianfan satellites are listed in Table 1 and their distribution is plotted in Figure 3. The mean (M) of these *regular* magnitudes is 5.75, while its standard deviation (SD) is 0.95 and the standard deviation of the mean (SDM) is 0.05. Those same quantities for the *orbit-raising* spacecraft are almost identical. M is only 0.02 fainter, SD is 0.02 smaller and SDM is 0.01 smaller.

Table 1. Magnitude Statistics

```
Apparent    Mean    SD    SDM
--------    ----    ----  ----
Regular     5.75    0.95  0.05
Raising     5.77    0.93  0.04
Averaged    5.76    0.94  0.04

1000-km     Mean    SD    SDM
--------    ----    ----  ----
Regular     5.23    0.70  0.04
Raising     5.25    0.75  0.03
Averaged    5.24    0.72  0.04
```

Apparent magnitudes can be standardized to a distance of 1,000 km by applying the inverse square law of light. This adjustment removes the observational selection bias due to varying satellite ranges and it also allows different models of satellites to be compared.

The 1000-km statistics in Table 1 list *regular* magnitude statistics of M = 5.23, SD = 0.70 and SDM = 0.04. The corresponding values for *orbit-raising* satellites are very similar, like the case of apparent magnitudes. The variational quantities are smaller for 1000-km magnitudes because the effect of differing ranges has been removed.

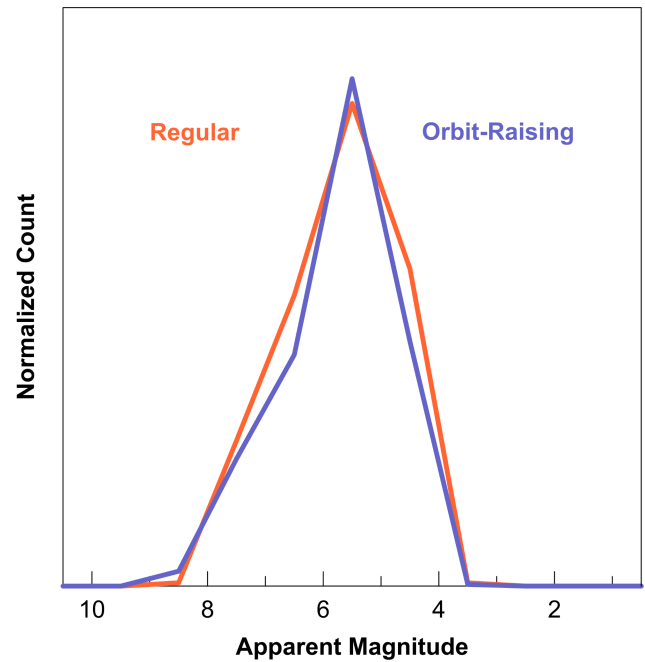

*Figure 3. The distribution of apparent magnitudes is nearly the same for regular and orbit-raising satellites.*

### 5. Tumbling satellites

Observations recorded by s2a systems, MMT9 and two visual observers indicate that some of the satellites from Launch 2 are tumbling. Figure 4 is the lightcurve of NORAD 61566 obtained by MMT9 during an 86-second track on 2025 March 10. Inspection by eye suggests a pattern of 4 brightness surges which repeats after about 45 seconds. Meanwhile, Figure 5 plots instrumental magnitudes from s2a systems for the same spacecraft as observed 11 days later. The different colored symbols represent 7 cycles of tumbling with a period of 45.14 seconds. The changing amplitude of brightness variations is likely due to spacecraft motion which resulted in shifting Sun-satellite-observer geometries during the track.

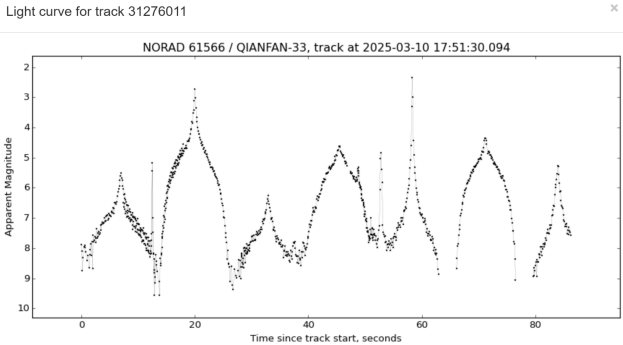

Figure 4. Lightcurve of NORAD 61566. Note the repeating pattern of brightness variations. Credit: MMT9 observatory.

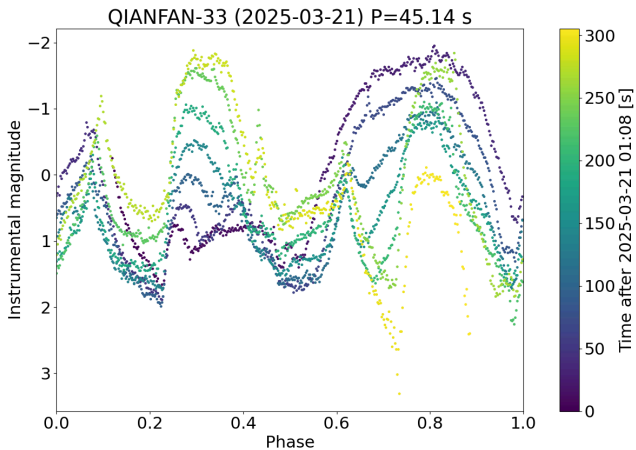

Figure 5. Lightcurve of NORAD 61566 showing 7 cycles of 45.11 second variations superposed. Credit: s2a systems.

Table 2 lists the six satellites that appear to be tumbling along with the dates of observation and the spin periods where they are known. The periods vary from about 13 to 170 seconds.

All six of the tumbling spacecraft are from Launch 2. As noted in Section 2, those satellites were made by a different manufacturer than the spacecraft in Qianfan launches 1, 3 and 4. McDowell indicates that all the tumbling spacecraft have failed.

Table 2. Tumbling satellites

| NORAD | Date       | Source      | Period* |
|-------|------------|-------------|---------|
| 61553 | 2025-03-18 | visual      | ?       |
| 61553 | 2025-03-20 | s2a systems | 13.28   |
| 61553 | 2025-03-27 | visual      | ?       |
| 61553 | 2025-04-18 | s2a systems | 21      |
| 61554 | 2025-03-21 | s2a systems | 91.04   |
| 61554 | 2025-04-18 | s2a systems | 170     |
| 61556 | 2025-03-20 | s2a systems | 55.33   |
| 61557 | 2025-03-10 | MMT9        | ?       |
| 61557 | 2025-03-12 | MMT9        | ?       |
| 61557 | 2025-04-12 | s2a systems | 19      |
| 61566 | 2025-03-10 | MMT9        | ~45     |
| 61566 | 2025-03-21 | s2a systems | 45.14   |
| 61566 | 2025-03-27 | visual      | ?       |
| 61567 | 2025-03-18 | visual      | ?       |
| 61567 | 2025-03-20 | s2a systems | 50.00   |
| 61567 | 2025-03-07 | MMT9        | ?       |

*seconds

**6. Physical model**

The techniques of generating simple physical brightness models were discussed by Mallama et al. (2024) which is referred to in this section as Paper 1. In these models the satellite is represented by a small number of surfaces and the reflection from these surfaces calculated using the position of the Sun and satellite with respect to the observer. The size and orientation of the surfaces and the type of reflection from those surfaces (for example, diffuse or specular) is adjusted depending on our knowledge of a particular spacecraft's design.

The conclusions of the work reported in Paper 1 on the first Qianfan launch were as follows:

1. The magnitude of the objects was well-modeled by diffusely reflecting, Earth-facing surfaces.
2. The objects were slightly brighter than the simple model when observed close to the horizon and close to the Sun azimuth. This behavior is often seen on other spacecraft in similar positions and can be attributed to forward-scattering of sunlight from Earth-facing surfaces on spacecraft.

A much larger dataset of 990 observations of objects from launches 1, 3 and 4 was available for the analysis presented here. The objects from launch 2 were not included in this dataset given they may be of a different design and also subject to the failures described earlier in this paper.

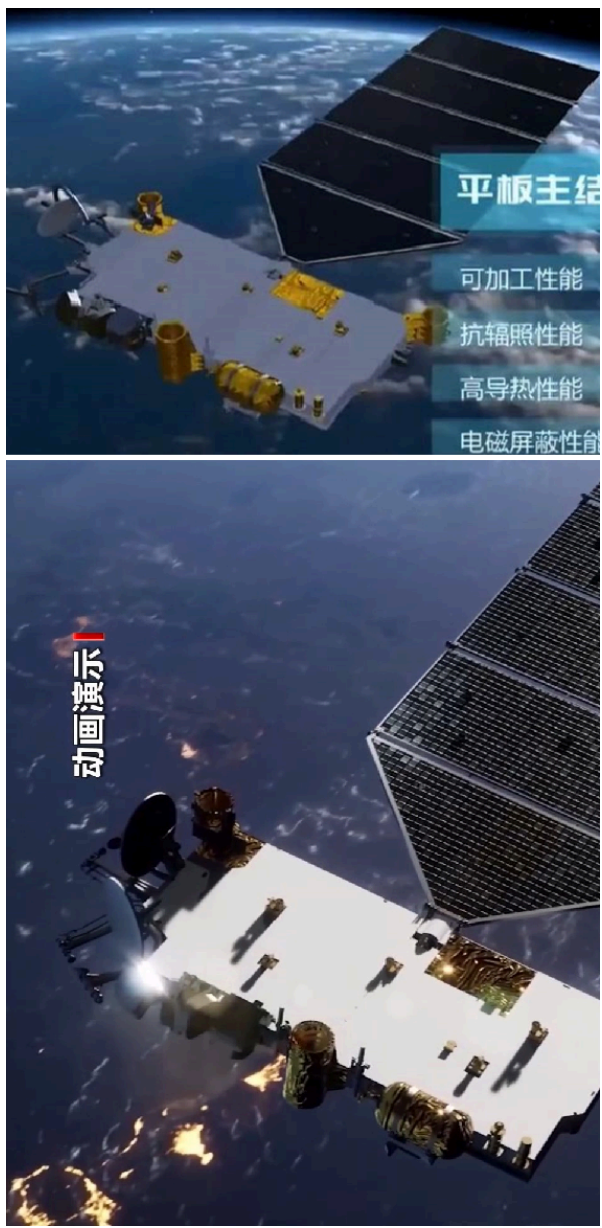

Figure 6. Two published images of a Qianfan spacecraft of the type believed used on launches 1, 3 and 4. In operations the solar-array would be rotated to track the Sun. Credit: https://www.youtube.com/watch?v=IG9VDSi8990

The goal of the model is to account for satellite brightness using a physical representation of the spacecraft and the reflecting properties of its surfaces.

The same model as used in Paper 1 was applied to the larger dataset, with the addition of a simple Bidirectional Reflectance Distribution Function (BRDF) to allow for the observed range of angles of reflection from the Earth-facing surface (0-50°). The BRDF was fitted as the cosine of the angle of reflection to the power of 1.7, Lambertian reflection from a perfectly diffuse surface would use a cosine power of one, as shown in Figure 7.

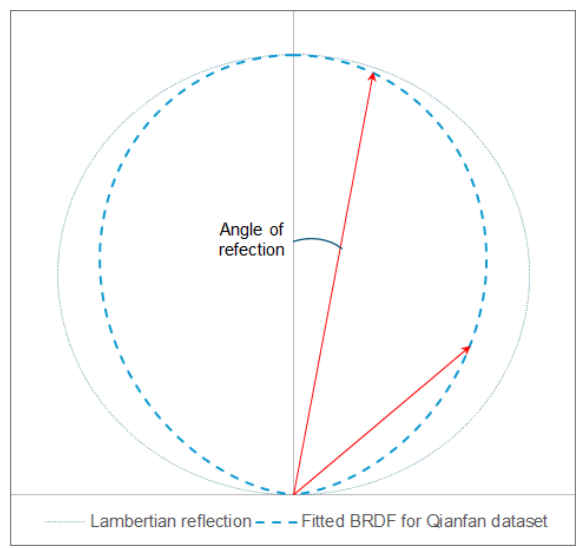

Figure 7. The Bidirectional Reflectance Distribution Function used in fitting the Qianfan data, compared to ideal Lambertian function.

Using this BRDF, the errors of the model are plotted against phase angle in Figure 8. For phase angles between 40° and 130° the errors are within the range of ±0.5 mag.

There are a number of observations at higher phase angles where the model underpredicts the brightness, as reported in Paper 1.

The larger dataset also shows a trend of underprediction of brightness at phase angles less than 60°, though not for all observations. This suggests that a Sun-azimuth-facing surface is sometimes visible at low phase angles, most probably the spacecraft solar-panel. Given that only a few observations show this effect, the visibility of the panel from the ground may be mitigated by restricting the elevation angle of the panel. SpaceX reported this technique was used on the Generation 1 Starlink spacecraft.

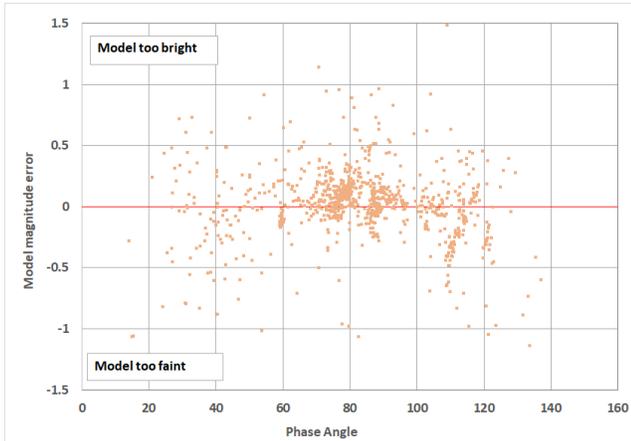

Figure 8. Model errors plotted against phase angle.

Figure 9 shows that the model remains accurate across a range of four observed magnitudes.

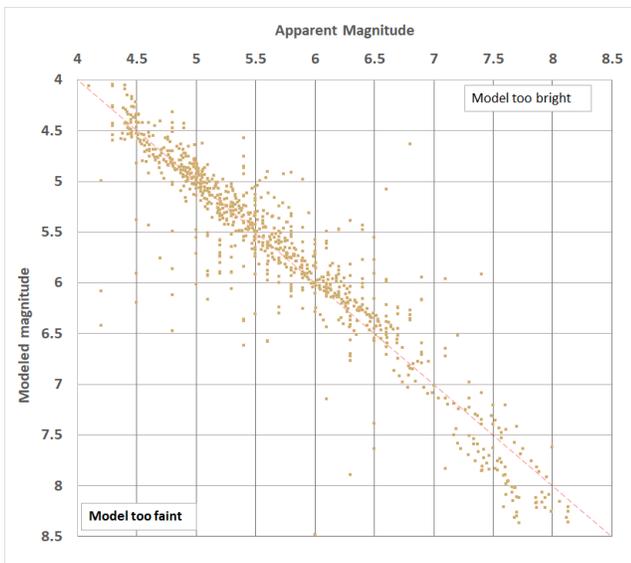

Figure 9. Model predicted magnitude plotted against the observed magnitude, the diagonal line shows a 1:1 correspondence of model and observations.

## 7. Impact on astronomy

Bright satellites cause streaks on astronomical images which compromise their scientific value. The International Astronomical Union (IAU, 2024) has established an "acceptable brightness limit" as indicated in Equations 1 and 2,

$V > 7.0$
for altitudes at or below 550 km

Equation 1

$V > 7.0 + 2.5 * \log_{10}( \text{altitude} / 550 )$
for altitudes above 550 km

Equation 2

where $V$ is the Johnson visual magnitude and *altitude* refers to the satellite's height above sea level.

At the lower altitude limit of 800 km for Qianfan satellites Equation 2 evaluates to magnitude 7.41, while at the upper limit of 1,070 km the magnitude is 7.72. Figure 10 shows that nearly all the magnitudes recorded for Qianfan satellites are brighter than the acceptable limit.

The limit described above applies to research performed with telescopes. Tyson et al. (2020) determined that streaks from brighter satellites cannot always be successfully removed from Rubin Observatory images.

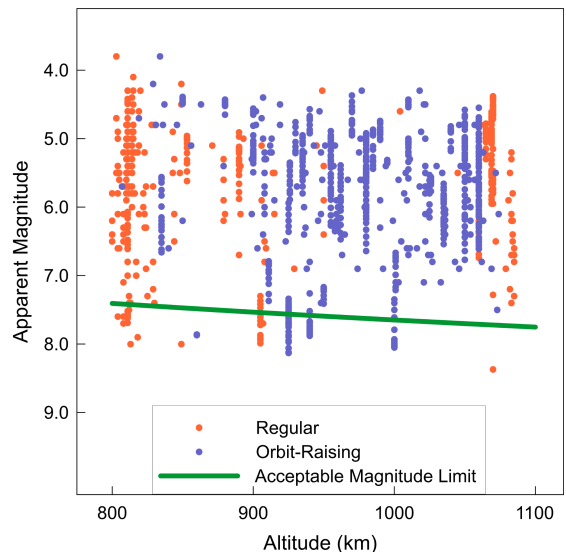

Figure 10. Observed magnitudes are compared to the acceptable brightness limit.

Meanwhile, satellites brighter than magnitude 6 spoil aesthetic appreciation of the night sky because they are visible to the unaided eye. The mean apparent magnitude for Qianfan satellites in Table 1 is 5.76, so most of them are brighter than that limit too.

Moreover, the analyses in this paper pertain to Qianfan satellites above 800 km altitude. Future satellites from this constellation are expected to orbit near 500 and 300 km like other telecommunication spacecraft. Bodies of the same dimensions and orientations would appear approximately 1 and 2 magnitudes brighter, respectively, at those heights.

## 8. Conclusions

The mean apparent and 1,000-km magnitudes of Qianfan spacecraft are 5.76 +/- 0.04 and 5.24 +/- 0.04, respectively. Several satellites from Launch 2 appear to be tumbling and they have probably failed. Nearly all of the observations can be modeled with diffusely reflecting, Earth-facing surfaces. These satellites will impact astronomical research and aesthetic appreciation of the night sky unless their brightness is mitigated.

## Acknowledgments

We thank the staff of the MMT9 observatory for maintaining a public database of their satellite observations. Two reviewers for the IAU/CPS offered comments that helped improve the manuscript. The Heavens-Above website was used to predict satellite passes. The planetarium program, Stellarium, was employed for data analysis. The Orbitron app was used for prediction and analysis.

## Appendix

Lightcurves of the tumbling satellites are illustrated below.

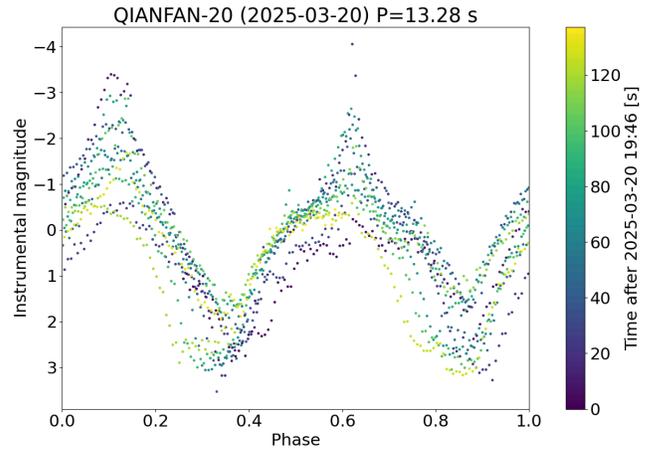

*Figure A-1. NORAD 61553. Credit: s2a systems.*

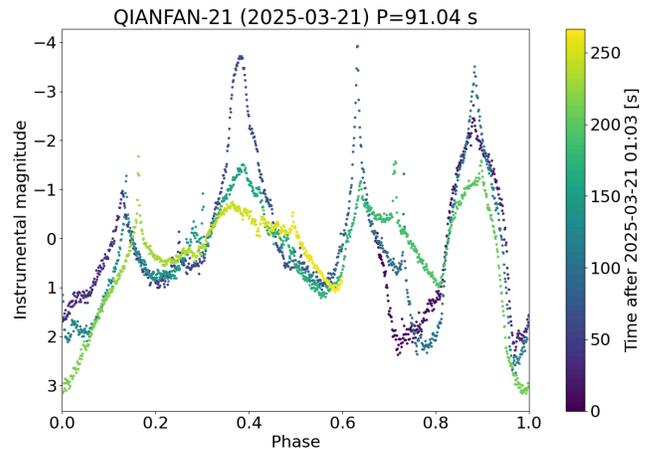

*Figure A-2. NORAD 61554. Credit: s2a systems.*

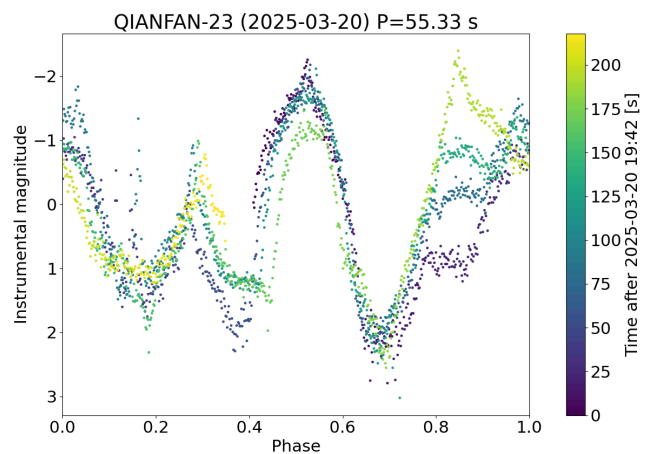

*Figure A-3. NORAD 61556. Credit: s2a systems.*

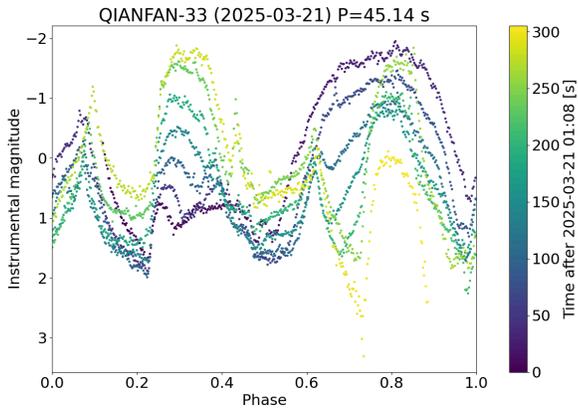

*Figure A-4. NORAD 61566. Credit: s2a-systems.*

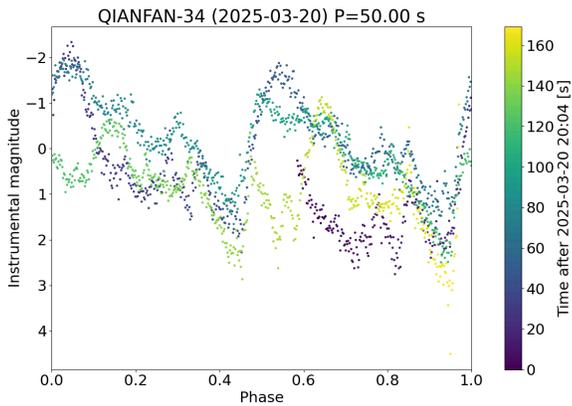

*Figure A-5. NORAD 61567. Credit: s2a systems.*

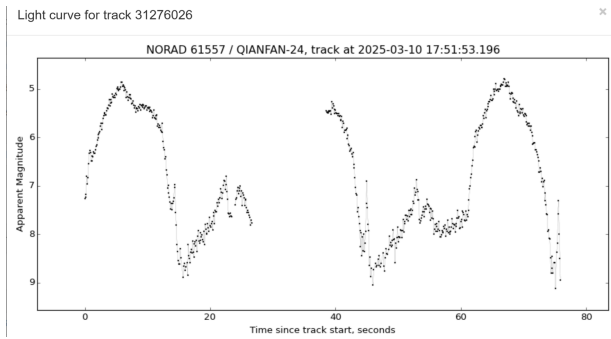

*Figure A-6. NORAD 61557. Credit: MMT9.*

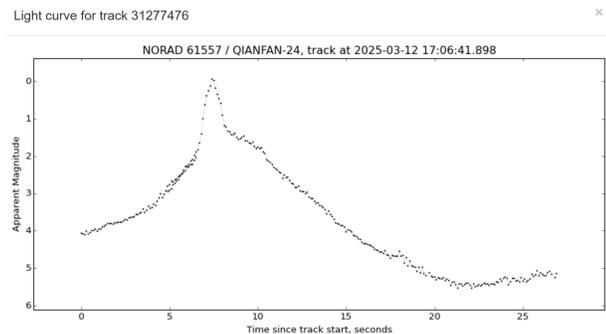

*Figure A-7. NORAD 61557. Credit: MMT9.*

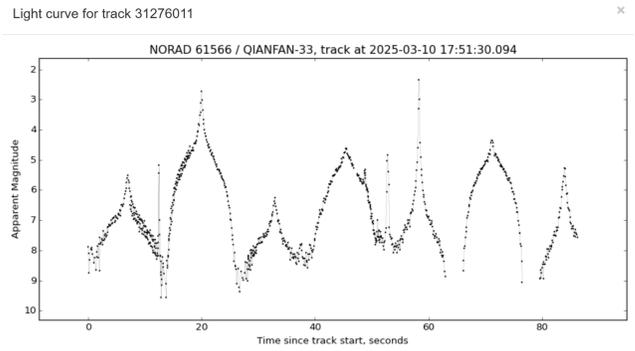

*Figure A-8. NORAD 61566. Credit: MMT9.*

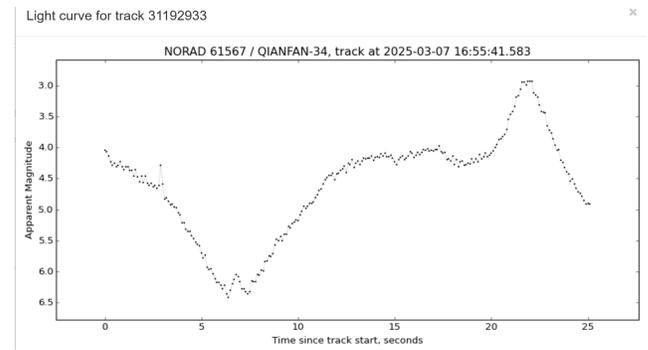

*Figure A-9. NORAD 61567. Credit: MMT9.*